\documentclass[a4paper]{jpconf}
\usepackage{graphicx}

\newcommand{\be}{\begin{equation}}
\newcommand{\ee}{\end{equation}}

\newcommand{\bea}{\begin{eqnarray}}
\newcommand{\eea}{\end{eqnarray}}

\begin{document}
\title{Reaction cross-section predictions for nucleon induced reactions}

\author{G. P. A. Nobre, I. J. Thompson, J. E. Escher
and F. S. Dietrich}

\address{Lawrence Livermore National Laboratory, P. O. Box 808, L-414, Livermore, CA 94551, USA}

\ead{nobre1@llnl.gov}

\begin{abstract}
A microscopic calculation of the optical potential for nucleon-nucleus scattering has been performed by explicitly coupling the elastic channel to all the particle-hole (p-h) excitation states in the target and to all relevant pickup channels. These p-h states may be regarded as doorway states through which the flux flows to more complicated configurations, and to long-lived compound nucleus resonances. We calculated the reaction cross sections for the nucleon induced reactions on the targets $^{40,48}$Ca, $^{58}$Ni, $^{90}$Zr and $^{144}$Sm using the QRPA description of target excitations, coupling to all inelastic open channels, and coupling to all transfer channels corresponding to the formation of a deuteron. The results of such calculations were compared to predictions of a well-established optical potential and with experimental data, reaching very good agreement. The inclusion of couplings to pickup channels were an important contribution to the absorption. For the first time, calculations of excitations account for all of the observed reaction cross-sections, at least for incident energies above 10 MeV.
\end{abstract}

\section{Introduction}
A quantitative description of nucleon-nucleus reactions is crucial for a broad variety of applications, including astrophysics, nuclear energy, 
radiobiology and space science \cite{FrontiersNuclSci2007,Townsend2002ASR30}. A fully microscopic description of such reactions is quite complex and resource-consuming, as one needs to consider not only the desired outcome
in an exit channel, but also the interference and competition processes for all other possible outcomes. 
A successful account of elastic nucleon-nucleus scattering, for example, has to include the effects from the excitation of non-elastic 
degrees of freedom, such as collective and particle-hole (p-h) excitations, transfer reactions, etc. 
Formally, all these non-elastic
effects can be accounted for by the projection-operator approach of Feshbach \cite{Feshbach1958ARNS8}.  Within the coupled reaction channels (CRC) framework, the flux removed from the elastic channel due to couplings to the non-elastic degrees of freedom can be obtained. 
An optical potential can therefore be defined \cite{Feshbach1958ARNS8,Brown1959RMP31} as the effective interaction in a 
single-channel calculation that contains
the effects of all the other processes that occur during collisions between nuclei. An important contribution to the optical potential arises directly from the calculation of  reaction cross section.
In the present work, we predict reaction cross-sections from Hartree-Fock-Bogoliubov (HFB) mean-field structure theory using ground state correlations corresponding to random phase approximation (RPA) and quasi-particle RPA (QRPA) descriptions of the excited states of the nuclei, based on energy density functional (EDF) theory, calculating the corresponding transition densities and transition potentials.

This work is part of the reaction section of the UNEDF (Universal Nuclear Energy Density Functional) project from SciDAC \cite{UNEDF,NobrePRL}. The main goal of this part of UNEDF is develop modern reaction theory based on microscopic nuclear structure input. The coupling between structure and reactions is accomplished by (i) using structural Hamiltonian matrix elements directly in microscopic calculations for scattering; (ii) using QRPA occupation amplitudes for ground and excited states to calculate DWBA matrix elements for neutron emission; (iii) adapting the density functional itself to give effective interactions between continuum states; (iv) using QRPA amplitudes to give the transition potentials for a full coupled-channels (CCh) calculation of scattering; (v) to extract the optical potential, and finally to (vi) to examine the statistical methods needed to average over compound-nucleus resonances for the optical potentials.

\section{Method}

Starting from self-consistent mean field calculation for the ground state, we obtain the initially  
occupied proton and neutron levels in a nucleus, using the Skryme energy-density functional SLy4 \cite[Table 1]{Chabanat1998NPA635}. This is a parametrization designed to  describe systems with arbitrary neutron excess, from stable to neutron matter, by improving isotopic properties, which overcomes deficiencies of other interactions away from the stability line.
A HFB calculation gives  the particle and hole levels of a given nucleus and fixes the p-h basis states for generating excited states within the framework of (Q)RPA, thus accounting for correlations caused by the residual interactions within the target.

Our scattering effective nucleon-nucleon interaction is of Gaussian shape, with parameters matched to  the volume integral and r.m.s. radius of the M3Y interaction at 40 MeV; it  includes a knock-on exchange correction \cite{Love}. 
In momentum space,  the central effective interaction is
$
v^{T}(q)= V_{0}^{T} (\pi /\mu_{T}^2)^{3/2} e^{-q^{2}/(2\mu_{T})^{2}},
$
with $V_0^{0} = -24.1921$ MeV and $\mu_{0} = 0.7180$ fm$^{-1}$ for the isoscalar part of the interaction and $V_0^{1} = 11.3221$ MeV and $\mu_{1} = 0.7036$ fm$^{-1}$ for the isovector component.
We do not include any imaginary part in this effective interaction, as our aim is to include all non-elastic excitations explicitly in our model, at least for scattering energies up to 40 MeV.
We convolute $ v^{T}(q)$ of the effective interaction with the transition densities to generate, after a reverse Fourier-Bessel transform,
the configuration space transition potentials.
The bare potential in the elastic channel is the single-folded potential using the ground-state density from the HFB calculation. For simplicity, this potential was also used for all excited states.

\section{Coupled reaction channels (CRC) calculations}

\begin{figure}[h]
\begin{minipage}{18pc}
\includegraphics[width=18pc]{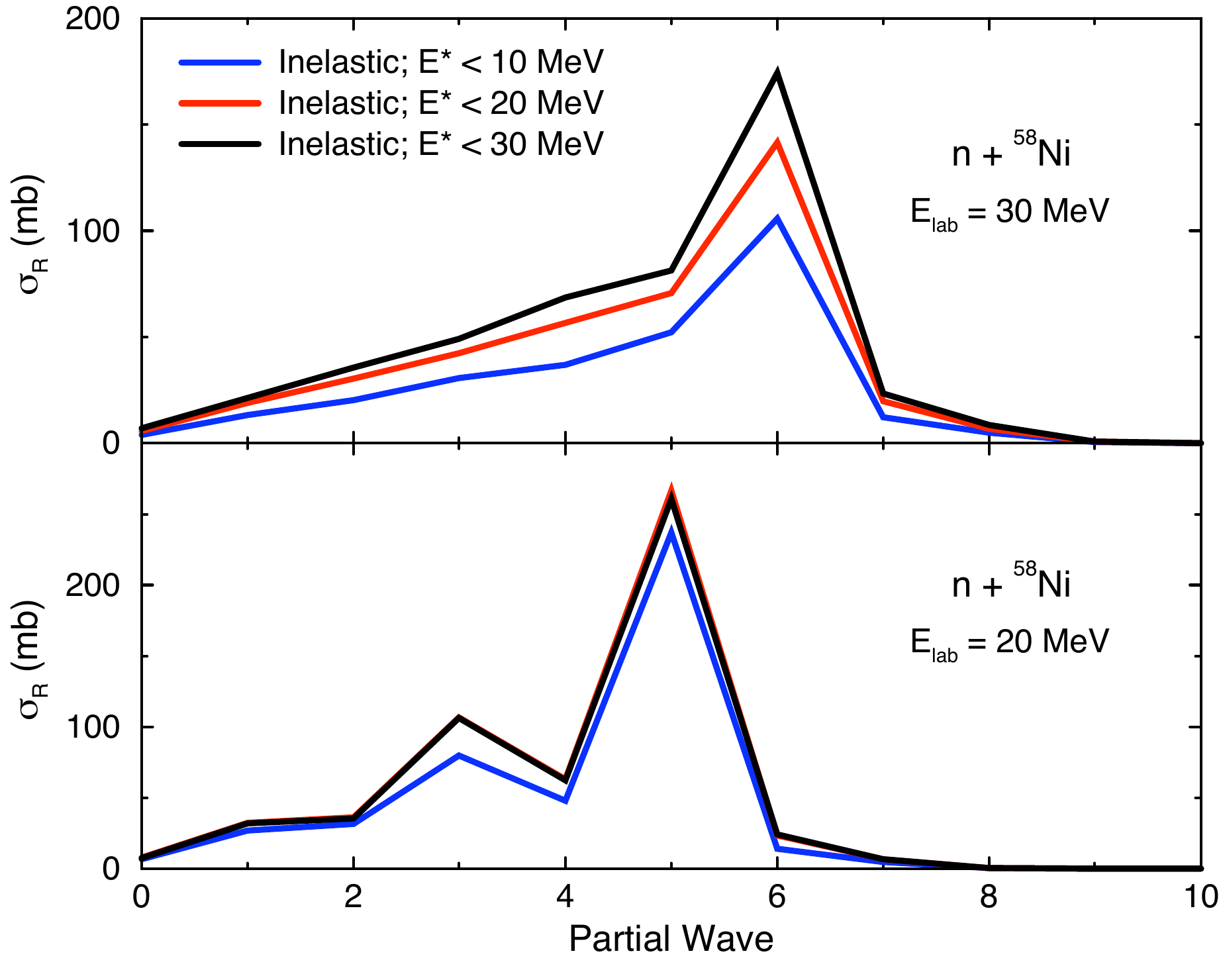}
\caption{\label{Fig:InelConv}Reaction cross-section as a function of the partial wave for the reaction n + $^{58}$Ni  at $E_{lab}$ = 20 MeV and 30 MeV.}
\end{minipage}\hspace{2pc}%
\begin{minipage}{18pc}
\includegraphics[width=18pc]{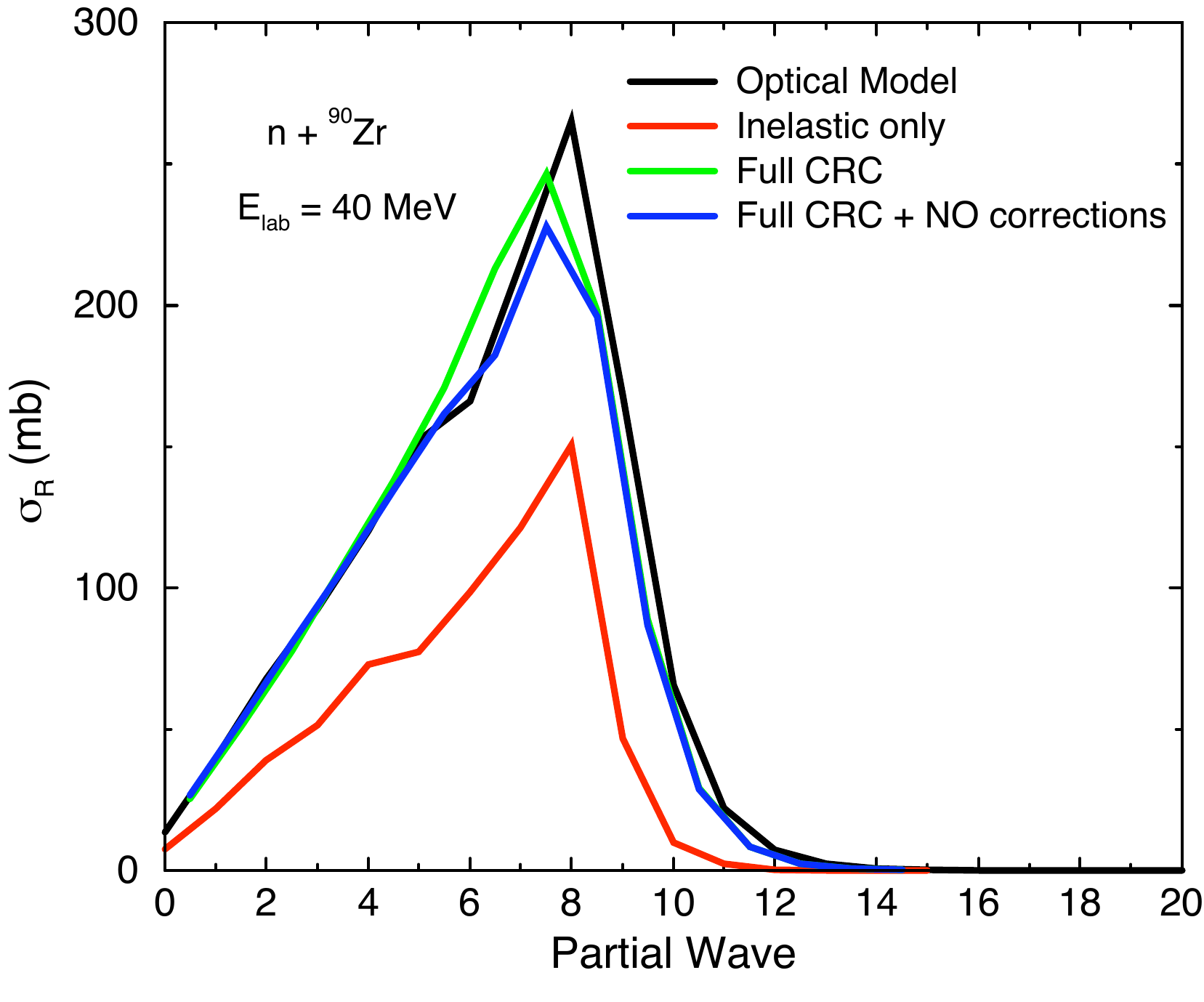}
\caption{Reaction cross-section as a function of the partial wave for the reaction n + $^{90}$Zr  at $E_{lab}$ = 40 MeV.}
\label{Fig:nZr90InelTransf}
\end{minipage} 
\end{figure}

We performed coupled channels calculations for reactions involving protons and neutrons scattered by the nuclei $^{40}$Ca, $^{48}$Ca, $^{58}$Ni, $^{90}$Zr and $^{144}$Sm, coupling the ground state to all levels with excitation energy ($E^{*}$) lying below some limit according to the QRPA model.
We examined the convergence with respect to maximum excitation energy, and found that convergence of the inelastic calculations requires coupling of  all excited levels below the scattering energy (i.e.\ all open channels).
To illustrate this behavior, we show in Figure~\ref{Fig:InelConv} the reaction cross sections for n + $^{58}$Ni obtained from calculations performed with couplings to all the excited states below 10, 20 and 30 MeV.
For  $E_{lab}$ = 20 MeV (Figure~\ref{Fig:InelConv}, lower panel) it is observed that coupling to excited states above the incident energy (20 MeV) produces no additional increment to the reaction cross section.

\subsection{Coupling to Pick-up Channels}

Although  the reaction cross section increases with the number of coupled states to the limit where all open channels are coupled (in Fig.~\ref{Fig:InelConv}), Figure \ref{Fig:nZr90InelTransf} serves as an example to show that these inelastic couplings account only for a small fraction 
of the reaction cross sections obtained from phenomenological optical model ($\sigma_{\mathrm{R}}^{\mathrm{OM}}$) calculations \cite{NobrePRL}. It is known \cite{Mackintosh2007PRC76,Keeley2008PRC77,Coulter1977NPA293} that pick-up channels  play an important role in nucleon-nucleus scattering. 
We therefore  included in our calculations couplings to all the channels leading to the formation of a deuteron, picking up the appropriate nucleon from any of the occupied levels from the target. 
For transfers, we approximate the HFB target states by bound single-particle states in a Woods-Saxon potential,  with the radii fitted to reproduce the volume radii and Fermi energy obtained by the HFB calculations.
The volume diffuseness and spin-orbit parameters were taken from Koning-Delaroche optical potentials \cite{Koning2003NPA713} 
at $E_{\rm lab}=0$, with spin-orbit radii adjusted by the same factor used to fit the volume part to HFB radii.
Due to the numerical impracticability of calculating couplings to inelastic and transfer channels simultaneously, we used the transfers optical potentials that reproduced the absorption from inelastic couplings in the first calculations. 

In CRC calculations we need, in addition to the scattering potentials in the incoming channel, the scattering potential between the deuteron and the remaining target.
We adopted the Johnson-Soper \cite{Johnson1970PRC1} prescription as it includes the effects of deuteron breakup in adiabatic (sudden) approximation. According to this prescription, the deuteron potential is the sum of the individual  neutron and proton potentials with the target. 
For the real parts we used the diagonal transition potentials  of the corresponding nucleon-nucleus reaction and,
for the imaginary parts, the sum of the  imaginary parts of the Koning-Delaroche  \cite{Koning2003NPA713}  optical potential for protons and neutrons on the target. That is, fitted parameters are used in the imaginary part of the deuteron potential, while we leave for future work to calculate deuteron and nuclear potentials self-consistently.

\subsection{Couplings between excited states}

\begin{figure}[h]
 \begin{center}
  \includegraphics[trim = 0mm 0mm 0mm 0mm, clip,scale=0.5,angle=-90.0]{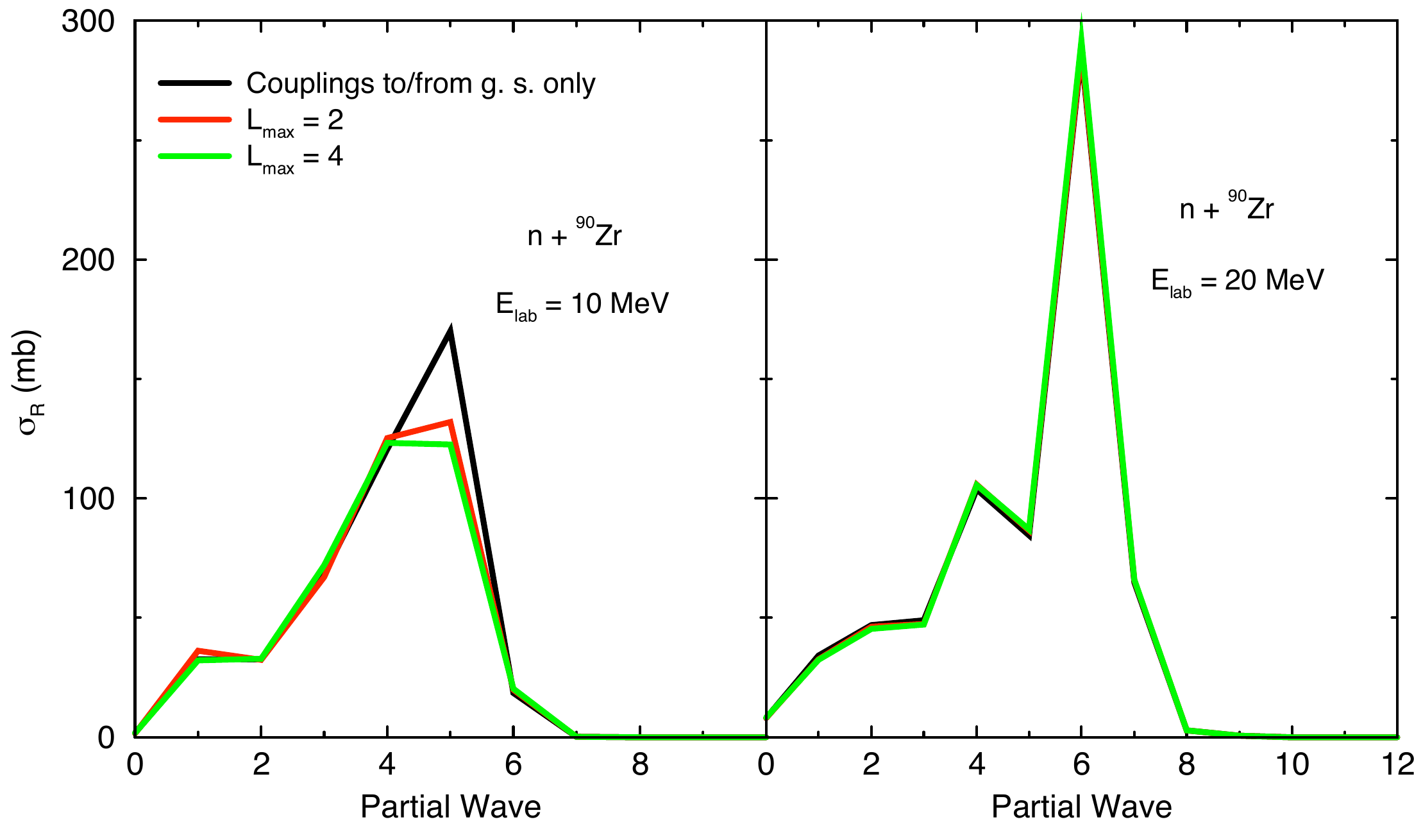} \\
 \end{center}
 \vspace{-4mm}
 \caption{Reaction cross-section as a function of the partial wave for the reaction n + $^{90}$Zr. Couplings to all RPA states below 20 MeV were included. The black lines represent calculations without couplings between excited states. The red and green lines correspond to the  results when transitions between excited states are explicitly calculated, with maximum transferred angular momentum $L_{max}=2$ and $L_{max}=4$, respectively.}
 \label{Fig:BetweenExcStates}
\end{figure}

Couplings \emph{between} excited states were explicitly considered as predicted by the RPA model, for nucleons scattered by $^{90}$Zr. In Figure~\ref{Fig:BetweenExcStates} we show the reaction cross-section as a function of partial wave for the reaction n + $^{90}$Zr at scattering energies of 10 MeV (left panel) and 20 MeV (right panel), where all RPA states lying below 20 MeV were coupled. For each energy we compare the calculations considering only couplings to and from the ground state with calculations that also include couplings between excited states (with a maximum value, $L_{max}$,  for the transferred angular momentum between them). It is observed that, although for $E_{lab}=10$ MeV there are small but noticeable differences between calculations, for higher energies the curves are almost undistinguishable. 
Despite the fact that, above 10 MeV, the couplings between excited levels do change the cross-sections of the individual states (mostly small changes, although for some few channels the cross-sections may differ by a factor up to 20\%), the overall sum among all states remains unchanged. This supports the validity of the \emph{doorway states} approximation, according to which it is not important, regarding reaction cross-sections, what processes occur after the excited states are populated from the ground state. Following this observation, couplings between excited states could be omitted from the CRC calculations.

\section{Results}

\begin{figure}[h]
\begin{minipage}[t]{18pc}
\includegraphics[width=18pc]{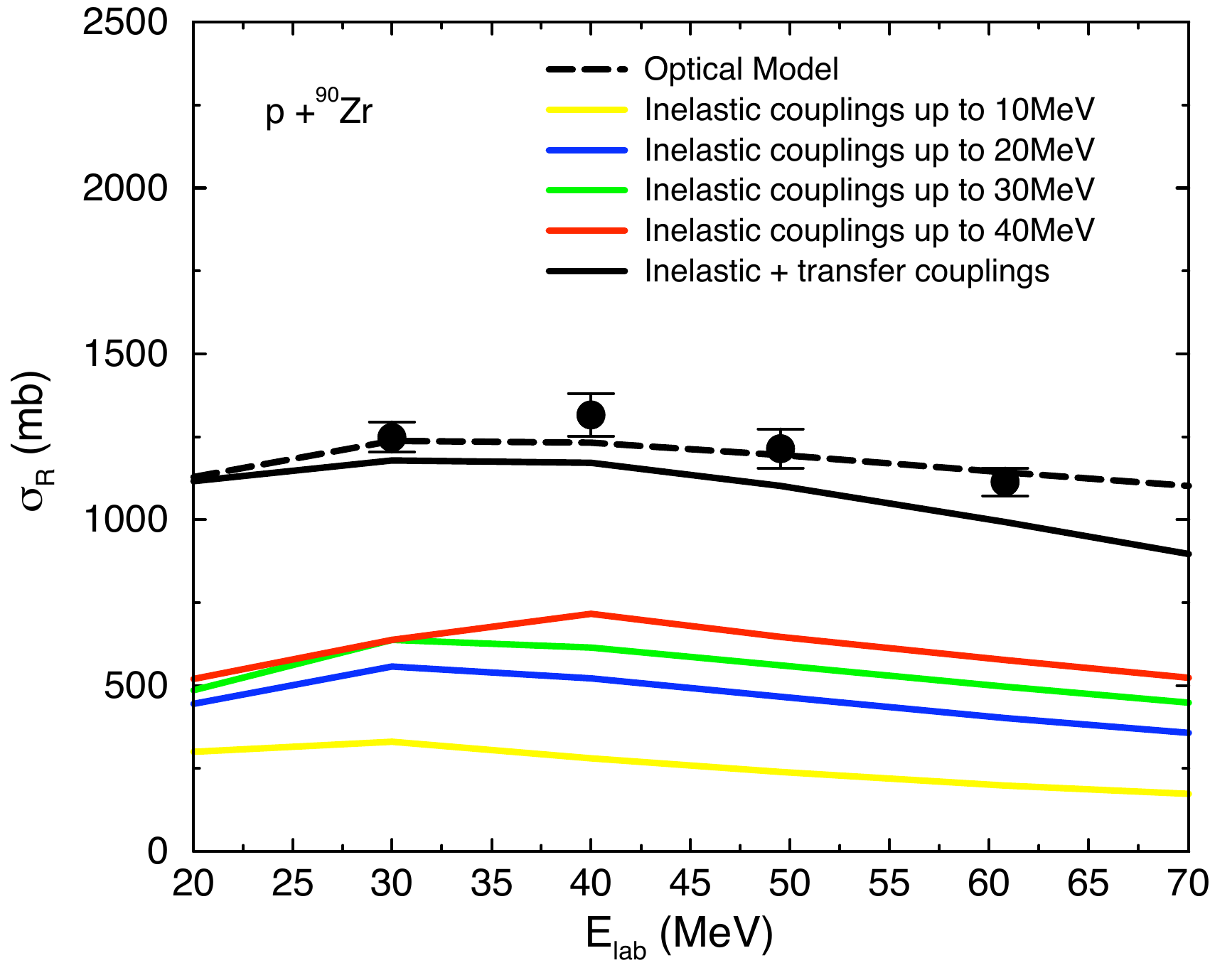}
\caption{Total reaction cross-section as a function of the incident energy for the reaction p + $^{90}$Zr.
 The Koning-Delaroche  \cite{Koning2003NPA713} optical model calculations are shown as dashed lines. The lines serve as guidance to the eye as calculations were performed only for $E_{\mathrm{lab}}$ = 10, 20, 30, 40, 50, 60 and 70 MeV. Data obtained from Refs. \cite{Menet1971PRC4}}
 \label{Fig:Zr90}
\end{minipage}\hspace{2pc}%
\begin{minipage}[t]{18pc}
\includegraphics[width=18pc]{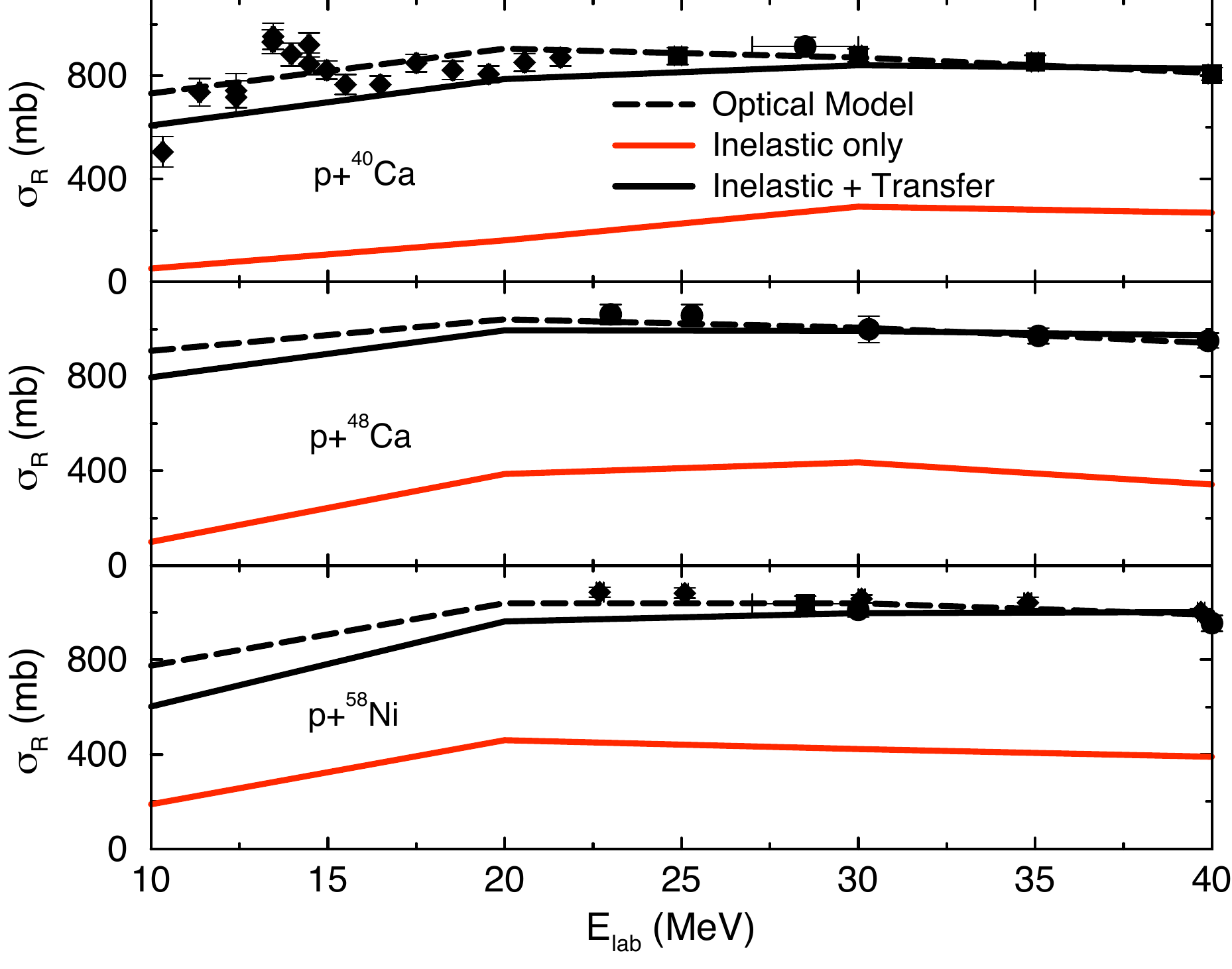}
 \caption{Total reaction cross-section as a function of the incident energy for the reactions of p + $^{40}$Ca,  p + $^{48}$Ca and  p + $^{58}$Ni. 
 The Koning-Delaroche  \cite{Koning2003NPA713} optical model calculations are shown as dashed lines. The lines serve as guidance to the eye as calculations were performed only for $E_{\mathrm{lab}}$ = 10, 20, 30 and 40 MeV. Data obtained from Refs. \cite{Turner1964NP58,Carlson1975PRC12,Dicello1970PRC2,Carlson1994PRC49,Menet1971PRC4,Eliyakut-Roshko1995PRC51}.}
 \label{Fig:SigRxElabData}
\end{minipage} 
\end{figure}

In Figure~\ref{Fig:Zr90} we show the predicted reaction cross sections for p + $^{90}$Zr obtained from different CRC calculations. It is clearly observed that the addition of inelastic couplings to higher excited states corresponds to an increase in the absorption until the limit where all open channels are coupled is reached.

After including couplings to the pickup channels through full CRC calculations, a large increment is found, approaching $\sigma_{\mathrm{R}}^{\mathrm{OM}}$ and the experimental data, as can be seen in Figures~\ref{Fig:Zr90} and \ref{Fig:SigRxElabData}.
 An even better agreement can be obtained after we include the non-orthogonality (NO) terms \cite[p. 226]{ThompsonBookNonOrthogonality} in the CRC calculations, as shown in Figure~\ref{Fig:nZr90InelTransf}. This correction arises because at small radii the deuteron bound state is not orthogonal to bound states occupied in the target.

\begin{figure}[h]
 \begin{center}
 \begin{minipage}[b]{19pc}
  \includegraphics[trim = 4mm 17mm 22mm 46mm, clip,width=19pc,angle=0.0]{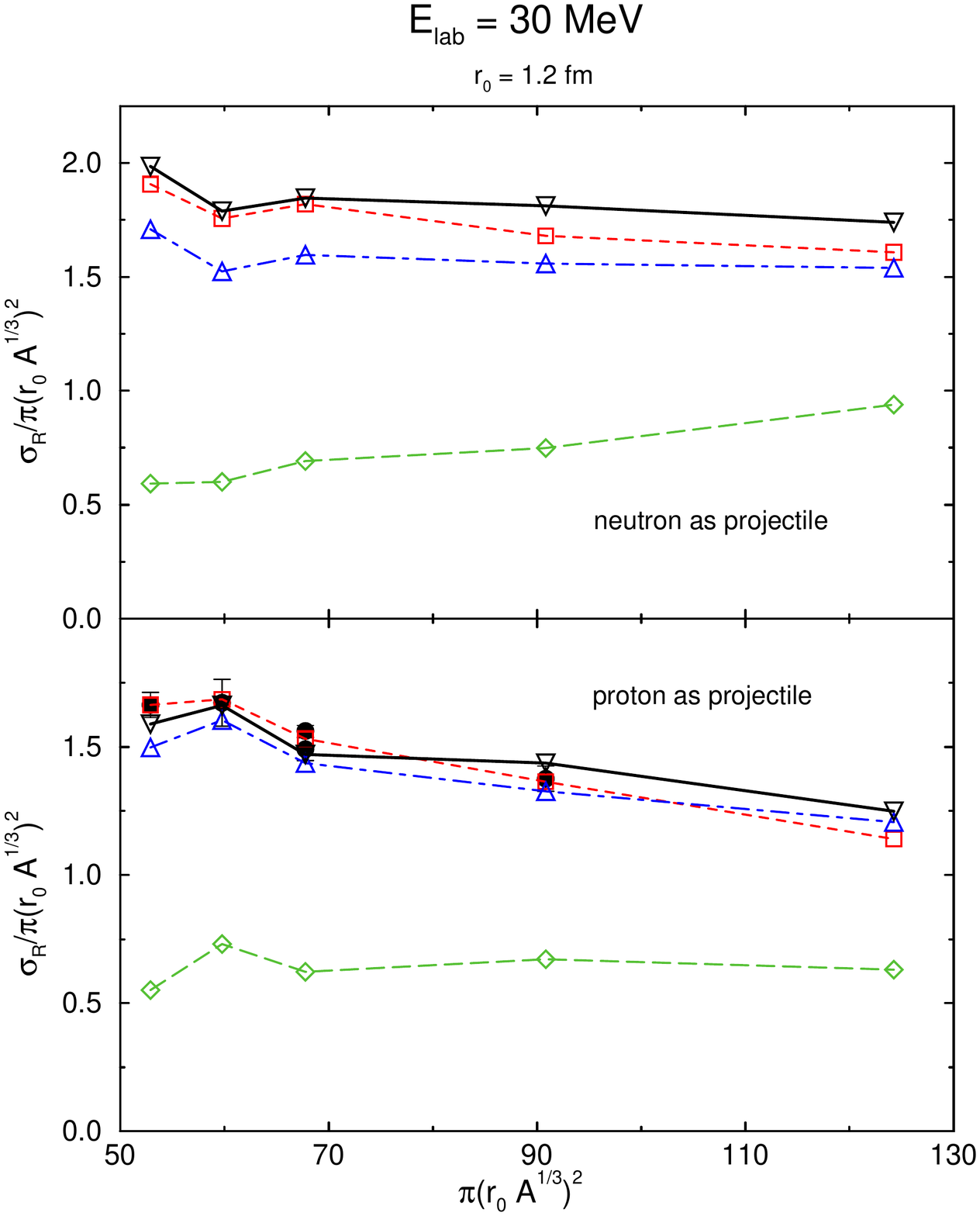} 
  \end{minipage}
  \hspace{2pc}
  \begin{minipage}[b]{15pc}
 \caption{Normalized total reaction cross-section at $E_{\rm lab}$ = 30 MeV  as a function of the area of the targets $^{40,48}$Ca, $^{58}$Ni, $^{90}$Zr and $^{144}$Sm. The value $r_{0}$ = 1.2 fm was used as a scale for the radii. The results are shown for couplings to the inelastic states lying below 30 MeV (dashed lines), to the inelastic states and to the transfer  channels (dash-dotted lines) and to the inelastic and transfer channels with non-orthogonality corrections (solid lines). The Koning-Delaroche  \cite{Koning2003NPA713} optical model calculations are shown as short-dashed lines. Open symbols are representing the calculations for the different target nuclei studied. Filled symbols are experimental data from Refs. \cite{Carlson1975PRC12,Carlson1994PRC49,Menet1971PRC4,Eliyakut-Roshko1995PRC51}}
 \label{Fig:allnucleiat30MeV}
  \end{minipage}
 \end{center}
\end{figure}

In Figure \ref{Fig:allnucleiat30MeV} we present the reaction cross sections obtained for nucleons scattered by the nuclei $^{40,48}$Ca, $^{58}$Ni, $^{90}$Zr and $^{144}$Sm at an incident energy of 30 MeV, as a function of the area  of the target. The absorption is shown relative to the reaction cross-section of a black sphere, which is approximately the geometrical area of the target. It can be seen again that, despite the important contribution of all inelastic couplings to the reaction cross-section, a large amount of absorption is due to the pickup channels and the treatment of the corresponding non-orthogonality corrections. Explicitly considering such couplings enabled us to account for practically all of doorway to the non-elastic cross-sections in the studied reactions.

\section{Conclusion}

In summary, we have calculated the reaction cross-sections for nucleon induced reactions  on nuclei    $^{40,48}$Ca, $^{58}$Ni, $^{90}$Zr and $^{144}$Sm, by explicitly calculating the couplings to all the doorway transfer and (Q)RPA inelastic  channels. We found that inelastic convergence is achieved when all open channels are coupled. 
While inelastic couplings account for an important part of the reaction cross section, most contributions come from couplings to the deuteron pickup channel,
in which case the non-orthogonality terms are significant. 
We obtain reaction cross sections that are in good agreement with phenomenological optical model results and experimental data. 
The importance of couplings between excited states was assessed, which led us to conclude that such couplings have no effect in the overall reaction cross sections, for incident energies above  10 MeV. This allowed us to consider only couplings to and from the ground state in the couple reaction channels calculations.
This work represents the first complete calculation that uses basic interactions between nucleons  within the nuclei to predict reaction observables for incident energy as low as 10 MeV, for which many-body approaches are no longer appropriate.
Using state-of-the-art nuclear structure models coupled with large-scale reaction computations allowed the accurate prediction of measurable quantities. This will serve as basis for future fully-consistent \emph{ab initio} developments for a range of nuclei including unstable species \cite{NobreBigPaper}.

\ack
 This work was performed under the auspices of the U.S. Department of Energy by Lawrence Livermore National Laboratory under Contract DE-AC52-07NA27344, and under SciDAC Contract DE-FC02-07ER41457.

\section*{References}
\bibliographystyle{unsrt} 
\bibliography{Proceeding-Nobre}

\begin{thebibliography}{10}

\bibitem{FrontiersNuclSci2007}
{NSAC 2007 Long Range Plan: \textit{The Frontiers of Nuclear Science}}.

\bibitem{Townsend2002ASR30}
{L.W. Townsend, F.A. Cucinotta and L.H. Heilbronn, \textit{Adv. Space Res.}
  {\bf 30}, 907 (2002)}.

\bibitem{Feshbach1958ARNS8}
{H. Feshbach, \textit{Annual Rev. of Nucl. Science} {\bf 8}, 49 (1958)}.

\bibitem{Brown1959RMP31}
{G.E. Brown, \textit{Rev. Mod. Phys.} {\bf 31}, 893 (1959)}.

\bibitem{UNEDF}
{UNEDF SciDAC Collaboration, \texttt{http://unedf.org}}.

\bibitem{NobrePRL}
{G. P. A. Nobre, F. S. Dietrich, J. E. Escher, I. J. Thompson, M. Dupuis, J.
  Terasaki, J. Engel, arXiv:1006.0267v1 [nucl-th], submitted to Phys. Rev.
  Lett.}

\bibitem{Chabanat1998NPA635}
{E. Chabanat, P. Bonche, P. Haensel, J. Meyer and R. Schaeffer, \textit{Nucl.
  Phys.} {\bf A635}, 231 (1998)}.

\bibitem{Love}
{G. Love, \textit{The (p,n) Reaction and the Nucleon-Nucleon Force}, in C.D.
  Goodman {\it et al.}, Plenum, 1980, from the \textit{Conference on the (p,n)
  Reaction and the Nucleon-Nucleon Force}, 1979, Telluride, Colorado}.

\bibitem{Mackintosh2007PRC76}
{R.S. Mackintosh and N. Keeley, \textit{Phys. Rev.} C {\bf 76}, 024601 (2007)}.

\bibitem{Keeley2008PRC77}
{N. Keeley and R.S. Mackintosh, \textit{Phys. Rev.} C {\bf 77}, 054603 (2008)}.

\bibitem{Coulter1977NPA293}
{P.W. Coulter and G.R. Satchler, \textit{Nucl. Phys.} {\bf A293}, 269 (1977)}.

\bibitem{Koning2003NPA713}
{A.J. Koning and J.P. Delaroche, \textit{Nucl. Phys.} {\bf A713}, 231 (2003)}.

\bibitem{Johnson1970PRC1}
{R.C. Johnson and P.J.R. Soper, \textit{Phys. Rev.} C {\bf 1}, 976 (1970)}.

\bibitem{Menet1971PRC4}
{J. J. H. Menet, E. E. Gross, J. J. Malanify and A. Zucker, \textit{Phys. Rev.}
  C {\bf 4}, 1114 (1971)}.

\bibitem{Turner1964NP58}
{J.F. Turner, B. W. Ridley, P. E. Cavanagh, G. A. Gard and A. G. Hardacre,
  \textit{Nucl. Phys.} {\bf 58}, 509 (1964)}.

\bibitem{Carlson1975PRC12}
{R.F. Carlson, A. J. Cox, J. R. Nimmo, N. E. Davison, S. A. Elbakr, J. L.
  Horton, A. Houdayer, A. M. Sourkes, W. T. H. van Oers and D. J. Margaziotis,
  \textit{Phys. Rev.} C {\bf 12}, 1167 (1975)}.

\bibitem{Dicello1970PRC2}
{J.F. Dicello and G. Igo, \textit{Phys. Rev.} C {\bf 2}, 488 (1970)}.

\bibitem{Carlson1994PRC49}
{R.F. Carlson, A. J. Cox, N. E. Davison, T. Eliyakut-Roshko, R. H. McCamis and
  W. T. H. van Oers, \textit{Phys. Rev.} C {\bf 49}, 3090 (1994)}.

\bibitem{Eliyakut-Roshko1995PRC51}
{T. Eliyakut-Roshko, R. H. McCamis, W. T. H. van Oers, R. F. Carlson and A. J.
  Cox, \textit{Phys. Rev.} C {\bf 51}, 1295 (1995)}.

\bibitem{ThompsonBookNonOrthogonality}
{I. J. Thompson and F. M. Nunes, \textit{Nuclear Reactions for Astrophysics},
  Cambridge U. Press (2009).}

\bibitem{NobreBigPaper}
{G. P. A. Nobre \emph{et al.}, \textit{in preparation}}.

\end{thebibliography}

\end{document}